\documentclass[preprints,perspective,accept,moreauthors,pdftex]{Definitions/mdpi}

\usepackage{amsmath} 
\newcommand{\angstrom}{\text{\normalfont\AA}}

\firstpage{1} 
\pubvolume{1}
\issuenum{1}
\articlenumber{0}
\pubyear{2021}
\copyrightyear{2020}
\datereceived{} 
\dateaccepted{} 
\datepublished{} 
\hreflink{https://doi.org/} 
	
\pdfoutput=1

\Title{Are Heavy Fermion Strange Metals Planckian?}

\TitleCitation{Are Heavy Fermion Strange Metals Planckian?}

\Author{Mathieu Taupin\orcidA{} and Silke Paschen\orcidB{}*}

\AuthorNames{Mathieu Taupin and Silke Paschen}

\AuthorCitation{Taupin, M.; Paschen, S.}

\address{%
Institute of Solid State Physics, Vienna University of Technology, Wiedner Hauptstr.\ 8-10, 1040 Vienna, Austria}

\corres{Correspondence: paschen@ifp.tuwien.ac.at; Tel.: +43-1-58801-13716}

\abstract{Strange metal behavior refers to a linear temperature dependence of the electrical resistivity at temperatures below the Mott-Ioffe-Regel limit. It is seen in numerous strongly correlated electron systems, from the heavy fermion compounds, via transition metal oxides and iron pnictides, to magic angle twisted bi-layer graphene, frequently in connection with unconventional or ``high temperature'' superconductivity. To achieve a unified understanding of these phenomena across the different materials classes is a central open problem in condensed matter physics. Tests whether the linear-in-temperature law might be dictated by Planckian dissipation---scattering with the rate $\sim k_{\rm B}T/\hbar$, are receiving considerable attention. Here we assess the situation for strange metal heavy fermion compounds. They allow to probe the regime of extreme correlation strength, with  effective mass or Fermi velocity renormalizations in excess of three orders of magnitude. Adopting the same procedure as done in previous studies, i.e., assuming a simple Drude conductivity with the above scattering rate, we find that for these strongly renormalized quasiparticles, scattering is much weaker than Planckian, implying that the linear temperature dependence should be due to other effects. We discuss implications of this finding and point to directions for further work.}


\begin{document}

\section{Introduction}
A first step in understanding matter is to delineate the different phases in
which it manifests. To do so, a characteristic that uniquely identifies a phase
must be found, and using its order has worked a long way. How this
classification should be extended to incorporate also topological phases
\cite{Sen17.1} is a matter of current research. Here, we focus on topologically
trivial matter and thus
take order parameter descriptions \cite{Lan37.1} as a starting point. As an
order parameter develops below a transition (or critical) temperature, the
system's symmetry is lowered (or broken). Cornerstones are the power law
behavior of physical properties near the critical temperature, with universal
critical exponents, and the associated scaling relationships. Combined with
renormalization-group ideas \cite{Wil75.1} this framework is now referred to as
Landau-Ginzburg-Wilson (LGW) paradigm. It has also been extended to zero
temperature. Here, phase transitions---now called {\em quantum} phase
transitions \cite{Sac11.1}---can occur as the balance between competing
interactions is tipped. To account for the inherently dynamical nature of the
$T=0$ case, a dynamical critical exponent needs to be added. This increases the
effective dimensionality of the system, which may then surpass the upper
critical dimension for the transition, so that the system behaves as
noninteracting of ``Gaussian''. Interestingly, however, cases have been
identified where this expectation is violated
\cite{Si01.1,Col01.1,Sen04.2,Sen04.3}, evidenced for instance by the observation
of dynamical scaling relationships \cite{Sch00.1} that should be absent
according to the above rationale. We will refer to this phenomenon as ``beyond
oder parameter'' quantum criticality. It appears to be governed by new degrees
of freedom specific to the quantum critical point (QCP). This is a topic of
broad interest both in condensed matter physics and beyond, but a general
framework is lacking. We will here discuss it from the perspective of heavy
fermion compounds, where it can manifests as Kondo destruction quantum
criticality \cite{Si01.1,Col01.1}. We will in particular discuss materials that
display linear-in-temperature ``strange metal'' electrical resistivity, as well
as the proposed relation \cite{Bru13.1,Leg19.1} to Planckian dissipation. We
will allude to similar phenomena in other materials platforms and point to directions for further research to advance the field.

\section{Simple models for strongly correlated electron systems}
Strongly correlated electron systems host electrons at the brink of
localization. The simplest model that can capture this physics is the Hubbard
model

\vspace{-0.1cm}
\begin{equation}
H = - t \sum_{\langle ij \rangle, \sigma} (d_{i\sigma}^{\dagger}d_{j\sigma}+d_{j\sigma}^{\dagger}d_{i\sigma}) + U \sum_i d_{i\uparrow}^{\dagger}d_{i\uparrow}d_{i\downarrow}^{\dagger}d_{i\downarrow}\; .
\label{Hubbard}
\end{equation}
The hopping integral $t$ transfers electrons from site to site and thus promotes itineracy, whereas the onsite Coulomb repulsion $U$ penalizes double occupancy of any site, thereby promoting localization. Thus, with increasing $U/t$, a (Mott) metal--insulator transition is expected. This simple model is suitable for materials where transport is dominated by one type of orbital with moderate nearest neighbor overlap, leading to one relatively narrow band. Well-known examples are found in transition metal oxides, for instance the cuprates. Here, the relevant orbitals are copper $d$ orbitals, kept at distance by oxygen atoms. The creation and annihilation operators are called $d$ and $d^{\dagger}$ here.

If two different types of orbitals interplay---one much more localized than the other---a better starting point for a theoretical description is the (periodic) Anderson model that, for the one-dimensional case, reads \cite{Hew97.1,Col15.1}

\vspace{-0.2cm}
\begin{equation}
H = \sum_{k,\sigma}\epsilon_k c^{\dagger}_{k\sigma}c_{k\sigma} 
+ \sum_{j,\sigma}\epsilon_f f^{\dagger}_{j\sigma}f_{j\sigma} + U\sum_{j}f^{\dagger}_{j\uparrow}f_{j\uparrow}f^{\dagger}_{j\downarrow}f_{j\downarrow} + \sum_{j,k,\sigma}V_{jk}(e^{ikx_j}f^{\dagger}_{j\sigma}c_{k\sigma} + e^{-ikx_j}c^{\dagger}_{k\sigma}f_{j\sigma}) \; .
\label{Anderson}
\end{equation}
Orbitals with large overlap, with the associated creation and annihilation operators $c$ and $c^{\dagger}$, form a conduction band with dispersion $\epsilon_k$. Orbitals with vanishing overlap are associated with the operators $f$ and $f^{\dagger}$. They are assumed to be separated by a distance greater than the $f$ orbital diameter and thus no hopping between them is considered. However, the hybridization term $V$ allows the $f$ electrons to interact. This model is particularly well suited for the heavy fermion compounds, which contain lanthanide (with partially filled $4f$ shells) or actinide elements (with partially filled $5f$ shells) in addition to $s$, $p$, and $d$ electrons. For close to integer filling of the $f$ orbitals, the Anderson model can be transformed into the Kondo (lattice) model

\vspace{-0.1cm}
\begin{equation}
H = \sum_{k,\sigma}\epsilon_k c^{\dagger}_{k\sigma}c_{k\sigma} 
- J \sum_{i} \vec{S}_i \cdot c^{\dagger}_{i,\sigma} \vec{\sigma}_{\sigma,\sigma'} c_{i\sigma'} \; ,
\label{Kondo}
\end{equation}
where the interaction between the localized and itinerant electrons is expressed in terms of an antiferromagnetic exchange coupling $J$. $\vec{S}$ is the local magnetic moment of the $f$ orbital and $\vec{\sigma}_{\sigma,\sigma'}$ are the Pauli spin matrices. One of the possible ground states of this model is a paramagnetic heavy Fermi liquid with a large Fermi surface, which contains both the local moment and the conduction electrons. The resonant elastic scattering at each site generates a renormalized band at the Fermi energy. Its width is of the order of the Kondo temperature $T_{\rm K}$, which can be orders of magnitude smaller than the noninteracting band width. In the (typically considered) simplest case (with a uniform and $k$ independent hybridization), this band extends across essentially the entire Brillouin zone.

In modern terms, this heavy fermion band could be seen as the realization of a nearly perfect ``flat band'', such as predicted \cite{Bis11.1} and later identified in magic angle twisted bi-layer graphene (MATBG) \cite{Cao18.2} as a result of moir\'e band formation, or expected in lattices with specific geometries \cite{Der15.1,Ley18.1} such as the kagome lattice \cite{Kan20.2,Ye21.1x} through destructive phase interference of certain hopping paths. Whereas the theoretical description of these flat band systems may be simpler than solving even the simplest Hamiltonians for strongly correlated electron systems, such as (\ref{Hubbard})-(\ref{Kondo}), the inverse might be true for the challenge on the experimental side. Heavy fermion compounds with a large variety of chemical compositions and structures \cite{Ste84.1,Ste01.1,Loe07.1} can be quite readily synthesized as high-quality (bulk) single crystals; the heavy fermion ``flat bands'' are robust (not fine tuned), naturally extend essentially across the entire Brillouin zone, and are pinned to the Fermi energy. Albeit, they form in the Kondo coherent ground state of the system, which is typically only fully developed at low temperatures. To realize such physics via a complementary route that might bring these properties to room temperature is an exciting perspective. Bringing together these different approaches bears enormous potential for progress. One aspect for which such crosstalk is already taking place is ``strange metal'' physics, which we address next.

\section{Strange metal phase diagrams}\label{strange}
Metals usually obey Fermi liquid theory, even in the limit of strong interactions. This is impressively demonstrated by the large body of heavy fermion compounds that, at sufficiently low temperatures, display the canonical Fermi liquid forms of the electronic specific heat

\vspace{-0.1cm}
\begin{equation}
C_p = \gamma T \; ,
\label{Cp_FL}
\end{equation}
the Pauli susceptibility

\vspace{-0.1cm}
\begin{equation}
\chi = \chi_0 \; ,
\label{chi_FL}
\end{equation}
and the electrical resistivity

\vspace{-0.1cm}
\begin{equation}
\rho = \rho_0 + A T^2\; ,
\label{rho_FL}
\end{equation}
where $\rho_0$ is the residual (elastic) resistivity. Theoretically, the
prefactors $\gamma$, $\chi_0$, and $A$ all depend on the renormalized electronic
density of states $N^*=N/N_0$, or the related renormalized (density-of-states)
effective mass $m^*=m/m_0 \sim N^*$, to first approximation as $\gamma \sim
m^*$, $\chi_0 \sim m^*$, and $A \sim (m^*)^2$. $N_0$ and $m_0$ are the free
electron quantities. Indeed, in double-logarithmic plots of  $\gamma$ vs
$\chi_0$ (Sommerfeld-Wilson) and $A$ vs $\gamma$ (Kadowaki-Woods), experimental
data of a large number of heavy fermion compounds fall on universal lines,
thereby confirming the theoretically expected universal ratios \cite{Kad86.1}.
The scaling works close to perfectly if corrections due to different ground
state degeneracies \cite{Tsu05.1} and effects of dimensionality, electron
density, and anisotropy \cite{Jac09.1} are taken into account.

The more surprising then was the discovery that this very robust Fermi liquid
behavior can nevertheless cease to exist. This can have multiple reasons, but
the predominant and best investigated one is quantum criticality
\cite{Sac99.1,Loe07.1,Pas21.1}. In the standard scenario for quantum criticality
of itinerant fermion systems \cite{Her76.1,Mil93.1,Mor12.2}, a continuously
vanishing Landau order parameter (typically of a density wave) governs the
physical properties. Its effect on the electrical resistivity is expected to be
modest because (i) the long-wavelength critical modes of the bosonic order
parameter can only cause small-angle scattering, which does not degrade current
efficiently, and (ii) critical density wave modes only scatter those areas on
the Fermi surface effectively that are connected by the ordering wavevector.
Fermions from the rest of the Fermi surface will short circuit these hot spots
\cite{Ros99.1}. For itinerant ferromagnets, $\rho \sim T^{5/3}$ is theoretically
predicted \cite{Sac99.1} and experimentally observed \cite{Smi08.1}. For
itinerant antiferromagnets, this type of order parameter quantum criticality
should result in $\rho \sim T^{\epsilon}$ with $1\leq \epsilon \leq 1.5$,
depending on the amount of disorder \cite{Ros99.1}. Whereas this may be
consistent with experiments on a few heavy fermion compounds, a strong
dependence of $\epsilon$ with the degree of disorder has, to the best of our
knowledge, not been reported.

Instead, a number of heavy fermion compounds exhibit a linear-in-temperature electrical resistivity

\begin{figure}[t!]
\vspace{-0.7cm}
\hspace{-4.5cm}\includegraphics[width=0.98\textwidth]{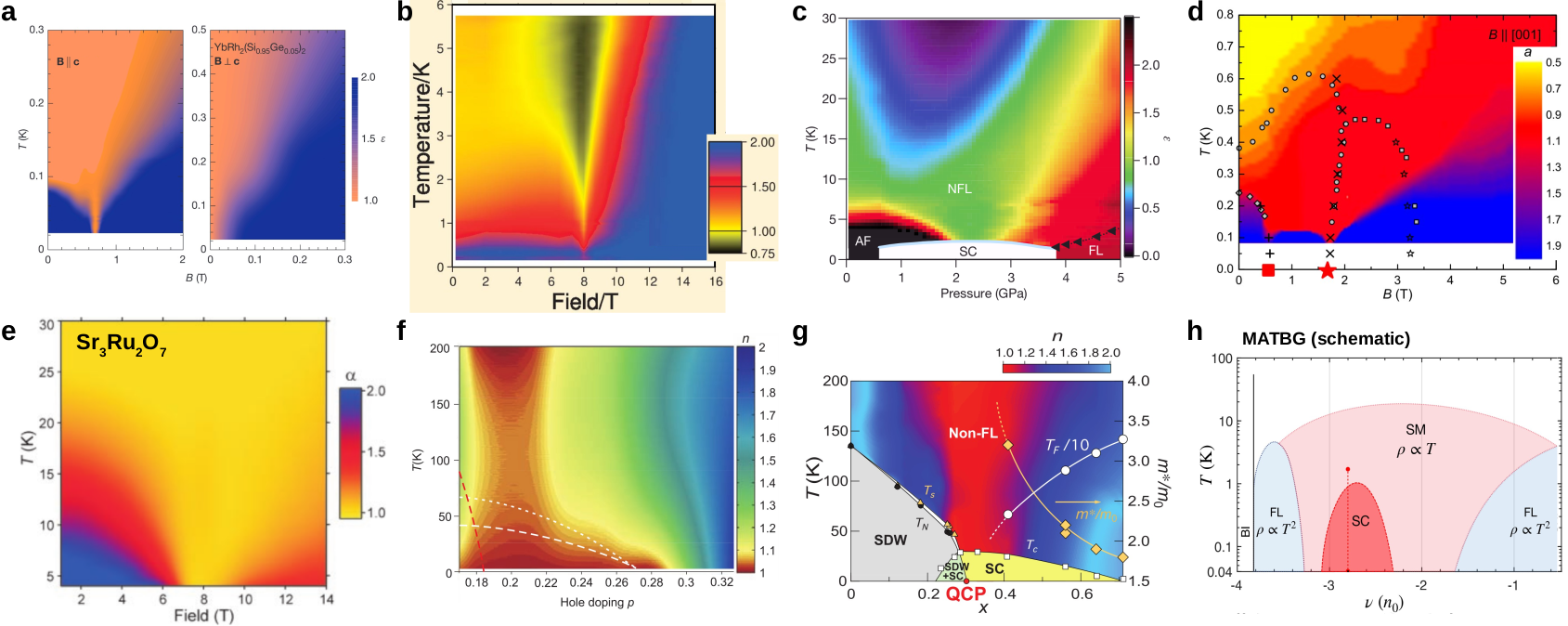}
\caption{Color-coded phase diagrams featuring strange metal behavior in various materials platforms. (\textbf{a}) YbRh$_2$Si$_2$ (left) and YbRh$_2$(Si$_{0.95}$Ge$_{0.05}$)$_2$ (right). From \cite{Cus03.1}. (\textbf{b}) CeRu$_2$Si$_2$. From \cite{Dao06.1}. (\textbf{c}) CeRhIn$_5$. From \cite{Par08.1}. (\textbf{d}) Ce$_3$Pd$_{20}$Si$_6$. From \cite{Mar19.1}. (\textbf{e}) SrRu$_3$O$_7$. Note that the temperature scale is cut at 4.5\,K. At lower temperatures, deviations from linear behavior towards larger powers are observed. From \cite{Gri01.2}. (\textbf{f}) La$_{2-x}$Sr$_x$CuO$_4$. From \cite{Coo09.1}. (\textbf{g}) BaFe$_2$(As$_{1-x}$P$_x$)$_2$. From \cite{Has12.1}. (\textbf{h}) Magic-angle twisted bi-layer graphene. From \cite{Jao21.1x}.
\label{fig1}}
\end{figure}

\vspace{-0.1cm}
\begin{equation}
\rho = \rho_0' + A' T \; ,
\label{rho_strange}
\end{equation}
a dependence dubbed ``strange metal'' behavior from the early days of high-temperature superconductivity on \cite{Nag92.1}. In Figure\;\ref{fig1}a-d we show four examples, in the form of temperature--magnetic field (a,b,d) or temperature--pressure (c) phase diagrams with color codings that reflect the exponent $\epsilon$ of the temperature-dependent inelastic electrical resistivity, $\Delta\rho \propto T^{\epsilon}$, determined locally as $\epsilon = \partial(\ln\Delta\rho)/\partial(\ln T)$. In all cases, fans of non-Fermi liquid behavior ($\epsilon \neq 2$) appear to emerge from QCPs, with $\epsilon$ close to 1 in the center of the fan and extending to the lowest accessed temperatures (at least in a,c,d). 

The most pronounced such behavior is found in YbRh$_2$Si$_2$
(Figure\;\ref{fig1}a,left). Below 65\,mK, the system orders
antiferromagnetically \cite{Tro00.2}. As magnetic field (applied along the
crystallographic $c$ axis) continuously suppresses the order to zero at 0.66\,T
\cite{Cus03.1}, linear-in-temperature resistivity, with $A' =
1.8\,\mu\Omega$cm/K and $\rho_0' = 2.43\,\mu\Omega$cm, extends from about 15\,K
\cite{Tro00.2} down to the lowest reached temperature (below 25\,mK)
\cite{Cus03.1}. Recently, this range was further extended down to 5\,mK, showing
$A' = 1.17\,\mu\Omega$cm/K for a higher-quality single crystal ($\rho_0' =
1.23\,\mu\Omega$cm) \cite{Ngu21.1}, thus spanning in total 3.5 orders of
magnitude in temperature. This happens in a background of Fermi liquid behavior
away from the QCP. A linear-in-temperature resistivity is also seen in the
substituted material YbRh$_2$(Si$_{0.95}$Ge$_{0.05}$)$_2$. Its residual
resistivity is about five times larger than that of the stoichiometric compound.
That this sizably enhanced disorder does not change the power $\epsilon$
indicates that the order-parameter description of an itinerant antiferromagnet
\cite{Ros99.1} is not appropriate here. This point will be further discussed in
Section\;\ref{FSjumps}.

For CeRu$_2$Si$_2$ (Figure\;\ref{fig1}b), the situation is somewhat more
ambiguous. Linear-in-temperature resistivity does not cover the entire core
region of the fan; both above 2\,K and below 0.5\,K, crossovers to other power
laws can be seen \cite{Dao06.1}. In CeRhIn$_5$ (Figure\;\ref{fig1}c), at the
critical pressure of 2.35\,GPa, linear-in-temperature resistivity extends from
about 15\,K down to 2.3\,K, the maximum critical temperature of a dome of
unconventional superconductivity \cite{Par08.1}. That the formation of emergent
phases such as unconventional superconductivity tends to be promoted by quantum
critical fluctuations is, of course, of great interest in its own right even if,
pragmatically, it can be seen as hindering the investigation of the strange
metal state. Finally, Ce$_3$Pd$_{20}$Si$_6$ exhibits two consecutive magnetic
field-induced QCPs, with linear-in-temperature resistivity emerging from both 
\cite{Mar19.1}. Other heavy fermion systems show similar behavior, though color-coded phase diagrams may not have been produced. A prominent example is CeCoIn$_5$. Its electrical resistivity was first broadly characterized as being linear-in-temperature below 20\,K down to the superconducting transition temperature of 2.3\,K \cite{Pet01.1}. Both magnetic field \cite{Pag03.1,Bia03.1} and pressure \cite{Sid02.1} suppress the linear-in-temperature dependence and stabilize Fermi liquid behavior, in agreement with temperature over magnetic field scaling of the magnetic Gr\"uneisen ratio indicating that a quantum critical point is situated at zero field \cite{Tok13.1}. Indeed, small Cd doping stabilizes an antiferromagnetic state \cite{Pha06.1}.

In Figure\;\ref{fig1}e-h we show resistivity-exponent color-coded phase diagrams of other classes of strongly correlated materials, a ruthenate, a cuprate, an iron pnictide, and a schematic phase diagram of MATBG. Also here, extended regions of linear-in-temperature resistivity are observed. Before we discuss this strange metal behavior in more detail in Section\;\ref{planckian}, we take a closer look at the Fermi liquid regions of the heavy fermion phase diagrams.

\section{Fermi liquid behavior near quantum critical points}\label{FL}
\begin{figure}[t!]
\hspace{0cm}\includegraphics[width=0.7\textwidth]{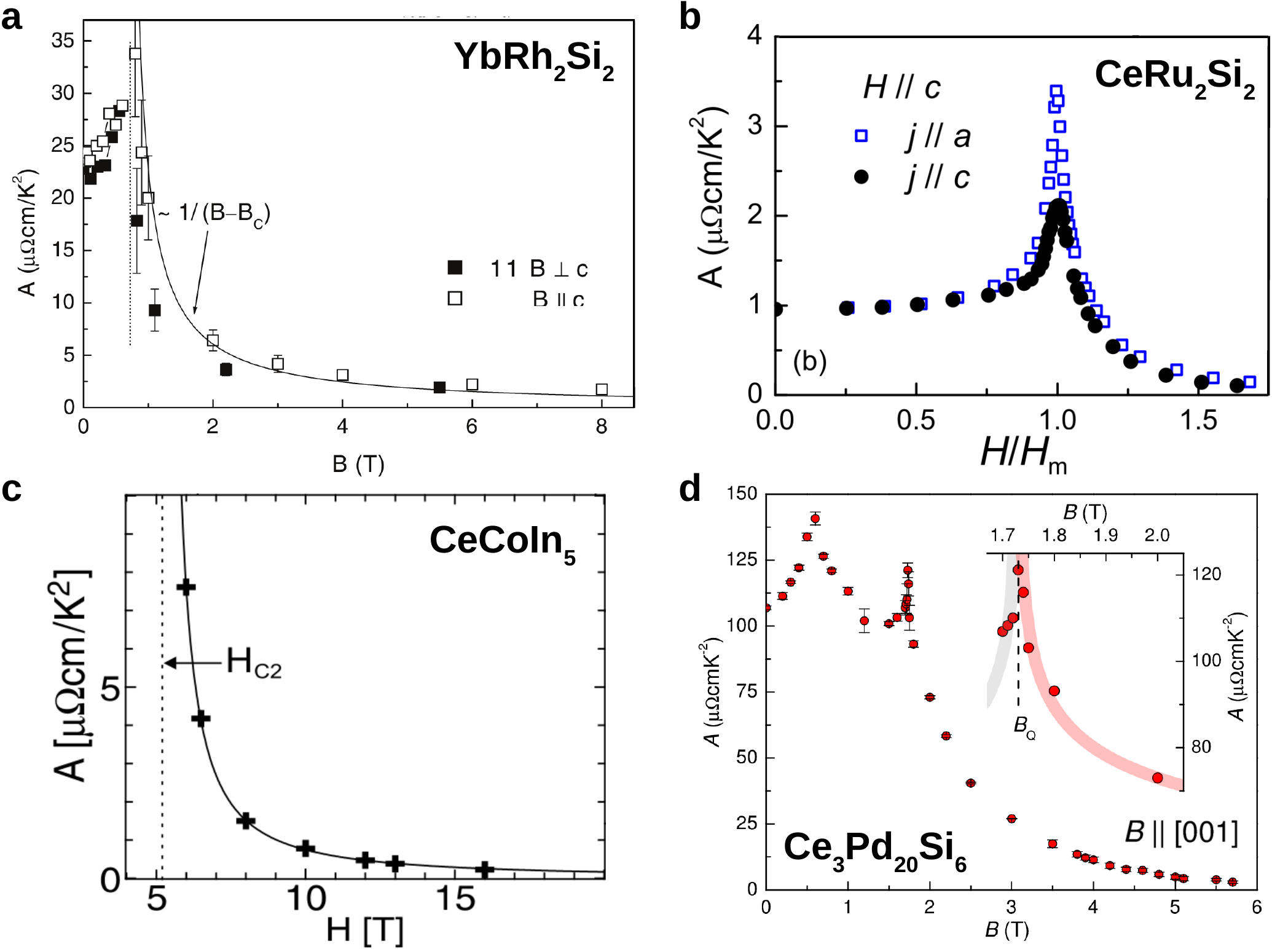}
\caption{Variation of the $A$ coefficient of the Fermi liquid form of the electrical resistivity, $\rho = \rho_0 + A T^2$, across QCPs in various heavy fermion compounds. (\textbf{a}) YbRh$_2$Si$_2$. From \cite{Cus03.1}. (\textbf{b}) CeRu$_2$Si$_2$. From \cite{Bou14.1}. (\textbf{c}) CeCoIn$_5$. From \cite{Pag03.1}. (\textbf{d}) Ce$_3$Pd$_{20}$Si$_6$. From \cite{Mar19.1}.
\label{fig2}}
\end{figure}

The low energy scales and associated low magnetic ordering temperatures
typically found in heavy fermion compounds call for investigations of these
materials at very low temperatures. Indeed, since early on, measurements down to
dilution refrigerator temperatures have been the standard. Because scattering
from phonons is strongly suppressed at such low temperatures, this is ideal to
study non-Fermi liquid  and Fermi liquid behavior alike. The phase diagrams in
Figure\;\ref{fig1}a-d all feature Fermi liquid regions, at least on the
paramagnetic side of the QCPs. The fan-like shape of the quantum critical
regions dictates that the upper bound of these regions shrinks upon approaching
the QCP. Nevertheless, high-resolution electrical resistivity measurements still
allow to extract the evolution of the Fermi liquid $A$ coefficient upon
approaching the QCP. In Figure\;\ref{fig2} we show such dependencies for four
different heavy fermion compounds. In all cases, the $A$ coefficient is very
strongly enhanced towards the QCP. In fact, within experimental uncertainty,
the data are even consistent with a divergence of $A$ at the QCP, as indicated
by the power law fits, $A \sim 1/(B - B_{\rm c})^a$, with $a$ close to 1, in
Figure\;\ref{fig2}a,c,d, suggesting that the effective mass diverges at the QCP.

This finding challenges the classification of heavy fermion compounds into lighter and heavier versions, that has been so popular in the early days of heavy fermion studies and that had culminated in the celebrated Kadowaki--Woods and Sommerfeld--Wilson plots. Which $A$ ($\gamma$, $\chi_0$) value should be used in these graphs? In \cite{Pas21.1} the use of lines instead of points in this graph was suggested, using actually measured values (and not fits to them) as end points. The question that remains is whether there is a ``background'' value, away from a quantum critical point, that is characteristic of a given compound. We will get back to this question in the next section.

\section{Strange metal behavior and Planckian dissipation}\label{planckian}
The occurrence of fans or, in some cases, differently shaped regions of linear-in-temperature resistivity in the phase diagrams of a broad range of correlated electron systems, as highlighted in Figure\;\ref{fig1}, raises the question whether a universal principle may be behind it. A frequently made argument is that linear-in-temperature resistivity is a natural consequence of the systems' energy scales vanishing at a quantum critical point and thus temperature becoming the only relevant scale. However, both the experimental observation of power laws $\Delta\rho\sim T^{\epsilon}$ with $\epsilon \neq 1$ in quantum critical heavy fermion compounds \cite{Mat98.1,Geg99.1,Kne01.1,Nak08.1} and predictions from order-parameter theories of such laws \cite{Ros99.1} tell us that this argument cannot hold in general. We thus have to be more specific and ask whether for quantum critical systems that {\em do} exhibit linear-in-temperature resistivities and, apparently, require description {\em beyond} this order parameter framework, a universal understanding can be achieved.

A direction that is attracting considerable attention
\cite{Bru13.1,Leg19.1,Gri21.1} is to test whether the transport scattering rate
$1/\tau$ of such systems may be dictated by temperature via

\begin{equation}
\frac{1}{\tau} = \alpha \frac{k_{\rm B}T}{\hbar}
\label{tau_planckian}
\end{equation}
with $\alpha \approx 1$. Should this be the case and $\tau$ be the only
temperature-dependent quantity in the electrical resistivity, then a
linear-in-temperature resistivity would follow naturally. Conceptually, this
roots in the insight, gained from the study of models without quasiparticles
\cite{Sac99.1,Zaa04.1,Sac11.1,Har16.1x,Har21.1x,Cho21.2x}, that a local
equilibration time (after the action of a local perturbation) of any many-body
quantum system cannot be faster than the Planckian time

\begin{equation}
\tau_{\rm P} = \frac{\hbar}{k_{\rm B}T}
\end{equation}
associated with the energy $k_{\rm B}T$ via the Heisenberg uncertainty principle \cite{Cho21.2x}. The question then is how to experimentally test this scenario. The simplest starting point is the Drude form for the electrical resistivity which, in the dc limit, reads

\begin{equation}
\rho = \frac{m}{n e^2} \frac{1}{\tau} \; ,
\label{Drude}
\end{equation}
with a temperature-independent effective mass $m$ and charge carrier concentration $n$, and (\ref{tau_planckian}) for the scattering rate $1/\tau$, leading to

\begin{equation}
\rho = \alpha \frac{m}{n e^2} \frac{k_{\rm B}T}{\hbar} \; .
\end{equation}
Interpreting this as the inelastic part of the linear-in-temperature electrical resistivity (\ref{rho_strange}), with $d\rho/dT = A'$, one obtains

\begin{equation}
\alpha = \frac{n}{m} \frac{e^2\hbar}{k_{\rm B}} A'
\label{alpha}
\end{equation}
or, in convenient units format,

\begin{equation}
\alpha = 2.15 \cdot \frac{n ({\rm nm}^{-3})}{m/m_0} \cdot A' (\mu\Omega{\rm cm/K})\; ,
\label{alpha_calc}
\end{equation}
where $m_0$ is the free electron mass. When this results in $\alpha \approx 1$,
the dissipation is said to be ``Planckian''. Before looking at experiments,
let's contemplate this for a moment. Relation (\ref{alpha}) is based on the
simple Drude model, and combines properties of well defined quasiparticles ($n$
and $m$) with a property that characterizes a non-Fermi liquid ($A'$)---possibly
one without quasiparticles---that is unlikely to follow the Drude model.
Furthermore, as shown in Section\;\ref{FL}, the Fermi liquid $A$ coefficient,
which is a measure of $m$, varies strongly with the distance to the QCP. Another
defining property of at least some of these strange metals are Fermi surface
jumps at the QCP (see Section\;\ref{FSjumps}). This adds a nontrivial
temperature and tuning parameter dependence to $n$. One should thus bear in mind
that choosing a simple Drude model as starting point holds numerous pitfalls. If
still doing so, it is unclear which $m$ and $n$ value to use.

In \cite{Bru13.1}, published quantum oscillation data, in part combined with results from density functional theory (DFT), were used to estimate $m$ and $n$ for a range of different materials, including also ``bad metals'' (see Section\;\ref{bad}) and simple metals in the regime where their resistivity is linear-in-temperature due to scattering from phonons. As an example, for Sr$_3$Ru$_2$O$_7$, de Haas--van Alphen (dHvA) data  \cite{Mer10.1} measured at dilution refrigerator temperatures on the low-field side of the strange metal fan (Figure\;\ref{fig1}e) were used. Contributions from the different bands, assumed as strictly 2D, were summed up as

\begin{equation}
\sigma = \tau \frac{e^2}{\hbar} \sum_i \frac{n_i}{m_i} \; ,
\label{sigma_sum}
\end{equation}
i.e., a constant relaxation time was assumed for all bands. Then, the heavy bands with small carrier concentration play only a minor role. In this way, $\alpha = 1.6$ was obtained. The dHvA effective masses of all bands were found to be modest (at most $10 m_0$) and essentially field-independent \cite{Mer10.1}, even though the $A$ coefficient increases by more than a factor of 8 on approaching the strange metal regime from the low field side \cite{Mer10.1}. The dHvA experiments may thus not have detected all mass enhancement \cite{Mer10.1,Bru13.1}. As shown below, using a larger effective mass would reduce $\alpha$.

\begin{figure}[t!]
\hspace{0cm}\includegraphics[width=0.7\textwidth]{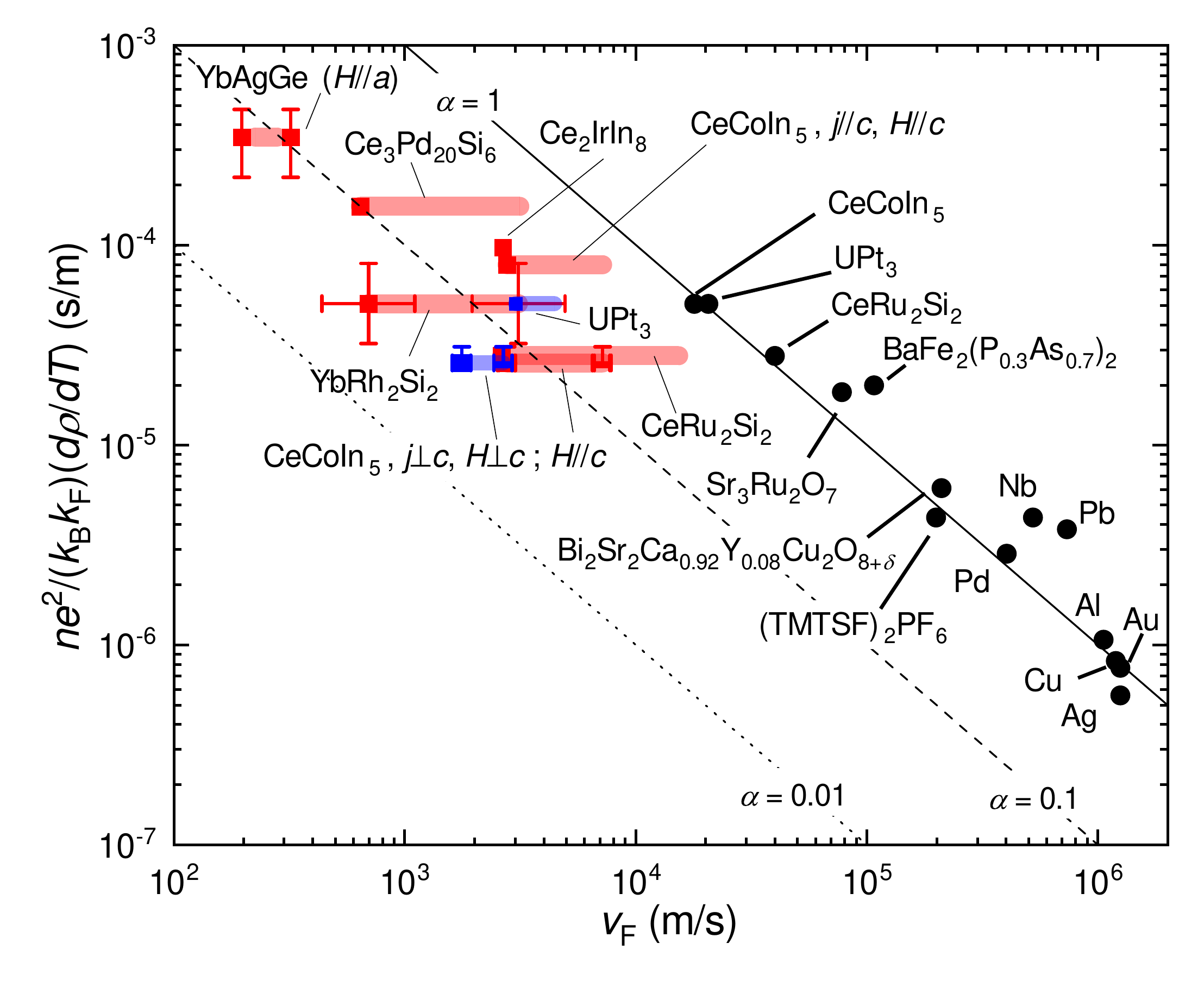}
\caption{Planckian dissipation plot of \cite{Bru13.1} revisited. Double-logarithmic plot of Fermi velocity $v_{\rm F}$ vs $ne^2/(k_{\rm B}k_{\rm F})(d\rho/dT) = \alpha/v_{\rm F}$ with data from \cite{Bru13.1} (black points) and data of the heavy fermion compounds listed in Table\;\ref{tab1} and analyzed here. The red squares result from the largest measured $A$ coefficient (or $\gamma$ value) for each compound near the strange metal regime, the shaded red lines from the published ranges of $A$ coefficient (or $\gamma$ value), and the error bars from uncertainties in the determination of the charge carrier concentration $n$ and sometimes other parameters (see Table\;\ref{tab1}). The full, dashed, and dotted line represent $\alpha = 1$, 0.1, and 0.01, respectively.
\label{fig3}}
\end{figure}

Similar analyses were performed for the other materials \cite{Bru13.1} and we replot the results as black points in Figure\;\ref{fig3}. The $x$ axis of this plot is the Fermi velocity $v_{\rm F}$ which, for a 3D system, can be brought into the form

\begin{equation}
v_{\rm F} {\rm (m/s)} = 3.58\cdot 10^{5} \cdot \frac{[n {\rm (nm}^{-3})]^{1/3}}{m/m_0} \; .
\label{v_F}
\end{equation}
The $y$ axis is the inverse of $v_{\rm F}$ multiplied by $\alpha$ (\ref{alpha_calc}) which, again for a 3D system, can be written as

\begin{equation}
\frac{\alpha}{v_{\rm F}} {\rm (s/m)} = 6.01\cdot 10^{-6} \cdot A' (\mu\Omega{\rm cm/K}) \cdot [n {\rm (nm}^{-3})]^{2/3} \; .
\label{v_F_alpha}
\end{equation}

To further assess how the results for $\alpha$ depend on the choice of the quasiparticle parameters $m$ and $n$, we here take a different approach. Instead of quantum oscillation data, we use global (effective) properties, namely the $A$ coefficient and the Hall coefficient $R_{\rm H}$, and estimate $\alpha$ for a number of strange metal heavy fermion compounds. Because of the extreme mass renormalizations observed in this class of materials (see Section\;\ref{FL}), it is particularly well suited for this test. Combining

\begin{equation}
\frac{m}{m_0}\cdot n^{1/3} = \frac{\gamma_{\rm mole-f.u.}}{V_{\rm f.u.}} \frac{3\hbar^2}{N_{\rm A} m_0 k_{\rm B}^2 (3\pi^2)^{1/3}}
\label{mass_gamma}
\end{equation}
with the Kadowaki--Woods ratio $A/\gamma^2 = 10^{-5}\,\mu\Omega$cm(mole K/mJ)$^2$, which is known to be very well obeyed in heavy fermion compounds \cite{Kad86.1}, we obtain

\begin{equation}
\frac{m}{m_0}\cdot [n ({\rm nm}^{-3})]^{1/3} = 3.26\cdot 10^4\frac{\sqrt{A (\mu\Omega{\rm cm}/{\rm K}^2)}}{V_{\rm f.u.} (\angstrom^3)}\; .
\label{mass_A}
\end{equation}
The rationale for using $A$ instead of $\gamma$ is that precise resistivity
measurements are most abundant in the literature (also under challenging
conditions such as high pressure and magnetic field) and that the resistivity is
much less sensitive to extra contributions from phase transitions than the
specific heat. 

A note is due on the determination of the charge carrier concentration $n$. It is commonly extracted from the Hall coefficient $R_\text{H}$, using the simple one-band relation $R_{\rm H}=1/ne$. Heavy fermion compounds are typically multiband systems, and thus compensation effects from electron and hole contributions can occur \cite{Fri10.1}. To limit the effect of anomalous Hall contributions, low-temperature data should be used \cite{Pas04.1}. Quantum oscillation experiments can determine the carrier concentration of single bands. However, heavy bands are hard to detect and it is unclear how to sum up contributions from different bands. An alternative is to determine $n$ via the superfluid density \cite{Orl79.1}, as was done previously \cite{Rau82.1,Ngu21.1}, using the relation (in cgs units)

\begin{equation}
n = \left( \frac{\xi_0 \cdot T_\text{c}\cdot \gamma}{7.95 \cdot 10^{-24}}\right)^{3/2} \; ,
\end{equation}
where $\xi_0$ is the superconducting coherence length, $T_{\rm c}$ the superconducting transition temperature, and $\gamma$ the normal-state Sommerfeld coefficient, which can be rewritten as

\begin{equation}
n (\text{nm}^{-3}) = 3020 \cdot \left( \frac{\xi_0 (\text{nm})\cdot T_\text{c} (\text{K})\cdot \gamma (\text{Jmol}^{-1}\text{K}^{-2})}{V_{\text{f.u.}} (\angstrom^3)}\right)^{3/2} \; .
\label{n_sc}
\end{equation}

Table\;\ref{tab1} lists the materials we inspected, with their $A$ coefficients
(or, when unavailable, $\gamma$), the best estimate of the charge carrier
concentration $n$ following the above discussion (see Table\;\ref{tab2} for
details), and the strange metal $A'$ coefficient. $m/m_0$ as calculated via
(\ref{mass_A}), or (\ref{mass_gamma}), is also listed.

\begin{specialtable}[b!] 
\small
\caption{Parameters used for Figures\;\ref{fig3} and \ref{fig4}. The red (or blue) square represents the largest $A$ coefficient (measured closest to the QCP), the shaded red (or blue) lines the range of $A$ coefficient measured upon moving away from the QCP. The Sommerfeld coefficient $\gamma$ is estimated from $A$ via the Kadowaki--Wood ratio, unless $A$ data are unavailable. The charge carrier concentrations $n$ and their error bars (where applicable) are taken from Table\,\ref{tab2}. For CeCoIn$_5$, several values are listed because the $A$ coefficient is different for in-plane ($H_a$) and out-of-plane ($H_c$) field, and the $A'$ coefficient is different for in-plane ($j_a$) and out-of-plane ($j_c$) currents. For YbAgGe, the $A'$ coefficient changes with field; the two extreme $A'$ values are denoted by the two red squares. For CeCoIn$_5$ ($j \perp c$), Figure\,\ref{fig3} shows the range $A'= (0.8 \pm 0.2)\,\mu\Omega$cm/K from \cite{Mak21.1x}. Data for Ce$_3$Pd$_{20}$Si$_6$ refer to the second QCP (near 2\,T, see Figure\;\ref{fig1}d) because for the lower field QCP no full data set on single crystals is published \cite{Mar19.1,Cus12.1}.}
\label{tab1}
\begin{tabular}{cccccc}
\toprule
\textbf{Compound} & $A$ ($\mu\Omega$cm/K$^2$) & $\gamma$ (J/molK$^2$) & $m/m_0$ & $n$ (nm$^3$)& $A'$ ($\mu\Omega$cm/K) \\
\midrule
Ce$_2$IrIn$_8$	&  - & 0.65 \cite{Kim04.1} & 183 & 2.5 & 8.8 \cite{Kim04.1} \\
Ce$_3$Pd$_{20}$Si$_6$ &  5 - 120 \cite{Mar19.1} & 0.707 - 3.46 & 136 - 665 & 1.7 & 18.3 \cite{Mar19.1} \\
CeCoIn$_5$ ($j_a, H_a$) &  12.4 - 28.3 \cite{Ron05.1} & 1.11 - 1.68 & 310 - 470 & 12.4 & 0.8 \cite{Tan07.1} \\
CeCoIn$_5$ ($j_a, H_c$) &  1.72 - 11.5 \cite{Ron05.1} & 0.414 - 1.07 & 116 - 300 & 12.4 & 0.8 \cite{Tan07.1} \\
CeCoIn$_5$ ($j_c, H_c$) &  1.72 - 11.5 \cite{Ron05.1} & 0.414 - 1.07 & 116 - 300 & 12.4 & 2.475 \cite{Tan07.1} \\
CeRu$_2$Si$_2$ &  0.1 - 3.4 \cite{Bou14.1} & 0.1 - 0.583 & 53 - 310 & 11.6 & 0.91 \cite{Dao06.1} \\
UPt$_3$ &  - & 0.425 - 0.625 \cite{Meu90.1} & 223 - 329 & 21.4 & 1.1 \cite{Bru13.1}\\
YbAgGe ($H//a$) &  - & 0.87 - 1.4 \cite{Tok06.1} & 1300 - 2100 & 1.6 & 27 - 59 \cite{Nik06.1} \\
YbRh$_2$Si$_2$ &  1.7 - 33.8 \cite{Geg02.1} & 0.41 - 1.85 & 250 - 1100 & 10 & 1.83 \cite{Geg02.1} \\
\bottomrule
\end{tabular}
\end{specialtable}

\begin{specialtable}[t!] 
\small
\caption{Charge carrier concentrations (in nm$^{-3}$) determined as follows: (i) $n_{\rm sc}$ from the superconducting coherence length $\xi_0$, the superconducting transition temperature $T_{\rm c}$, and the normal-state Sommerfeld coefficient $\gamma$, all in zero field, via (\ref{n_sc}); (ii) $n_{\rm H}$ from the Hall coefficient at the lowest temperatures, where anomalous contributions are minimal, via $R_{\rm H}=1/ne$; (iii) $n_{\rm qo}$ from quantum oscillation experiments reviewed in \cite{Bru13.1}, by summing up the carrier concentrations from all detected bands. For CeCoIn$_5$, the $\gamma$ coefficient is taken at 2.5~K, without taking into account the logarithmic divergence. The error bar in $n$ used for CeCoIn$_5$ ($j \perp c$) in Figure\;\ref{fig3} reflects the range of the parameters given in \cite{Mak21.1x}. YbRh$_2$Si$_2$ is close to being a compensated metal, resulting in a strong sensitivity of $n$ to small differences in the residual resistivity. The largest reported $R_{\rm H}$ value, which corresponds to $n_{\rm H}=26.0$ \cite{Pas04.1}, has the lowest compensation and is thus most accurate. Nevertheless, the $R_{\rm H}$ value of LuRh$_2$Si$_2$ is even larger, corresponding to $n_{\rm H} =  11.6$\,nm$^{-3}$ \cite{Fri10.1}, suggesting that there is still some degree of compensation in the sample of \cite{Pas04.1}. We list the average of both values, 18.8\,nm$^{-3}$, as best $n_{\rm H}$ estimate. For the plots, we use the approximate average of $n_{\rm sc}$ and
$n_{\rm H}$, i.e., 10\,nm$^{-3}$, with an asymmetric error bar $\delta n_+ = 10$\,nm$^{-3}$ and $\delta n_- = -5$\,nm$^{-3}$ (see Table\;\ref{tab1}). Similar compensation effects are also encountered in UPt$_3$ \cite{Kam99.2}. Bold fonts indicate the values used for the $\alpha$ estimates (see Table\,\ref{tab1}).}
\label{tab2}
\begin{tabular}{cccccccc}
\toprule
\textbf{Compound} & $\xi_0$ (nm) & $T_{\rm c}$ (K) & $\gamma$ (J/molK$^{2}$) & $n_{\rm sc}$ & $n_{\rm H}$ & $n_{\rm qo}$\\
\midrule
Ce$_2$IrIn$_8$	&  - & - & - & - & \textbf{2.5} \cite{Sak03.1} & - \\
Ce$_3$Pd$_{20}$Si$_6$ &  -  & - & - & - & \textbf{1.7} \cite{Mar19.1} & -  \\
CeCoIn$_5$ &  5.6 \cite{Zho13.1} & 2.3 \cite{Pet01.1} & 290 \cite{Pet01.1} & 10.8 & 10.1 \cite{Shi07.2}-12.5 \cite{Mak21.1x} & \textbf{12.4} \cite{Bru13.1} \\
CeRu$_2$Si$_2$ & - & - & - & - & 3.1 \cite{Dao06.1} -7.8 \cite{Had86.1} & \textbf{11.6} \cite{Bru13.1} \\
UPt$_3$ &  12 \cite{Che86.1} & 0.52 \cite{Che86.1} & 0.43 \cite{Meu90.1} & 22.4 & 9 \cite{Had86.1} & \textbf{21.4} \cite{Bru13.1} \\
YbAgGe & - & - & - & - & \textbf{1.6} \cite{Bud05.1} & - \\
YbRh$_2$Si$_2$ & 97 \cite{Ngu21.1} & 0.0079 \cite{Ngu21.1} & 1.42 \cite{Ngu21.1} & 4.86 & 18.8 \cite{Pas04.1,Fri10.1} & - \\
\bottomrule
\end{tabular}
\end{specialtable}

All these data are then included in Figure\;\ref{fig3} in the following way. The $v_{\rm F}$ (\ref{v_F}) and $\alpha/v_{\rm F}$ (\ref{v_F_alpha}) value resulting from the largest measured $A$ coefficient (or $\gamma$ value) for each compound is shown as red square. The shaded red lines represent the published ranges of $A$ coefficient (or $\gamma$ value). The error bars represent uncertainties in the determination of the charge carrier concentration (see Table\;\ref{tab1}). Lines for $\alpha = 1$, 0.1, and 0.01 are also shown. It is clear that none of the shaded red lines overlaps with the $\alpha = 1$ line. The discrepancy with the points extracted from quantum oscillation experiments \cite{Bru13.1} is quite striking.

\begin{figure}[t!]
\hspace{0cm}\includegraphics[width=0.7\textwidth]{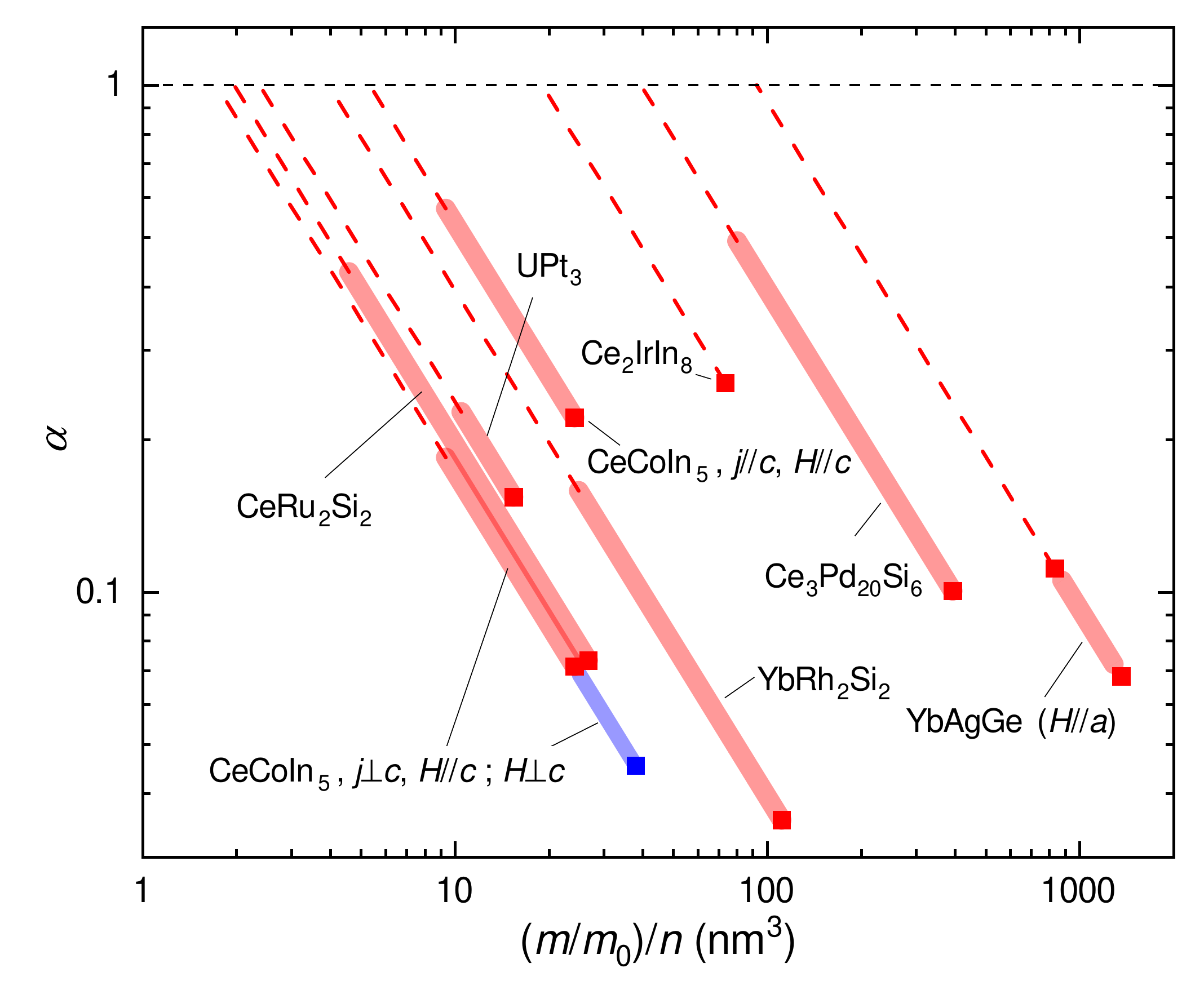}
\caption{No Planckian dissipation from heavy quasiparticles in heavy fermion compounds. Double-logarithmic plot of $\alpha$ vs $(m/m_0)/n$ for various strange metal heavy fermion compounds, as given in Table\;\ref{tab1}. Red squares and shaded lines have the same meaning as in Figure\;\ref{fig3}. The dashed lines are to help reading off the values of $(m/m_0)/n$ for which the linear-in-temperature electrical resistivity in these compounds could be governed by Planckian dissipation. Note that in all cases the ``Planckian dissipation'' effective masses obtained in this way are sizably smaller than even the smallest values experimentally accesses by tuning the systems away from the strange metal regime (top end of full shaded lines).
\label{fig4}}
\end{figure}

In Figure\;\ref{fig4} we present these results in a different form, as $\alpha$ vs $(m/m_0)/n$. The red squares and red shaded lines have the same meaning as in Figure\;\ref{fig3}. The dashed lines are extrapolations of the shaded lines to $\alpha=1$. We can thus directly read off the values of $(m/m_0)/n$ for which a given compound would, in this simple framework, be a Planckian scatterer. In all cases, this is for effective masses significantly smaller than even the smallest measured ones in the Fermi liquid regime.

What are the implications of this finding? We first comment on the discrepancy with the results from \cite{Bru13.1}. Apparently, averaging the contributions from different bands detected in quantum oscillation experiments via (\ref{sigma_sum}) leads to sizably larger Fermi velocities (sizably smaller effective masses) than our $A$ coefficient approach. In heavy fermion compounds, a coherent heavy fermion state forms at low temperatures, and the Fermi liquid $A$ coefficient is known to be a pertinent measure thereof. It is thus either the use of (\ref{sigma_sum}) that should be reconsidered or the reliance in quantum oscillation experiments to detect the heaviest quasiparticles. Clearly, if dissipation in strange metal heavy fermion compounds is to be Planckian, this would hold only for the very weakly renormalized quasiparticles---perhaps some form of ``background'' to renormalizations from quantum critical fluctuations.

\section{Linear-in-temperature resistivity beyond the strange metal regime}\label{bad}
In a number of strongly correlated electron systems, including quasi-2D conductors such as the high-$T_{\rm c}$ cuprates but also 3D transition metal oxides and alkali-doped fullerides, linear-in-temperature resistivity is observed beyond the Mott--Ioffe--Regel (MIR) limit \cite{Hus04.1,Gun03.1}. At this limit, the electron mean free path approaches the interatomic spacing. Semiclassical transport of long-lived quasiparticles should then cease to exist and the resistivity should saturate \cite{Iof60.2,Mot72.1}, in 3D systems of interest to us here to \cite{Wer17.1}

\begin{equation}
\rho_{\rm MIR} = \frac{h}{e^2}\cdot a  \; ,
\end{equation}
where $a$ is a relevant interatomic distance. Using the Drude resistivity (\ref{Drude}) with the Fermi velocity $v_{\rm F}=\hbar k_{\rm F}/m$, the Fermi wave vector $k_{\rm F}=(3 \pi^2 n)^{1/3}$, and the mean free path $\ell = \tau v_{\rm F}$ one obtains

\begin{equation}
\rho = \frac{h}{e^2}\cdot \frac{3\pi}{2} \frac{1}{k_{\rm F}^2 \ell} =  \frac{h}{e^2}\cdot a \cdot C \; ,
\end{equation}
where the value of the constant $C$ depends on details of the electronic and crystal structure. For the MIR limit, i.e.\ $C=1$, one gets

\begin{equation}
\rho_{\rm MIR} (\mu\Omega{\rm cm}) = 258 \cdot a (\angstrom)  \; ,
\label{rho_MIR}
\end{equation}
As pointed out in \cite{Cho21.2x}, linear-in-temperature resistivity beyond the MIR limit is possible even from electron-phonon scattering \cite{Wer16.1} and should be referred to as ``bad metal'' \cite{Eme95.1} as opposed to ``strange metal'' behavior.

Room temperature resistivities of heavy fermion compounds are typically much lower than this limit \cite{Hus04.1}. For YbRh$_2$Si$_2$,  $\rho(300\,{\rm K}) = 80$\,$\mu\Omega$cm \cite{Tro00.2}. Inserting the lattice parameters $a = 4.007$\,\AA\ and $c = 9.858$\,\AA\ \cite{Tro00.2} in (\ref{rho_MIR}) gives $\rho_{\rm MIR} \approx 1000\,\mu\Omega$cm and $\approx 2500\,\mu\Omega$cm, respectively. However, that the resistivity stays (well) below these values should not be seen as evidence for resistivity saturation in the MIR sense. Instead, the ``flattening'' of the temperature dependence of the electrical resistivity in heavy fermion compounds is known to result from the crossover between coherent Kondo scattering at low temperatures and incoherent Kondo scattering (with a negative temperature dependence) at high temperatures \cite{Hew97.1,Col15.1}. Only when, at high enough temperatures, electron-phonon scattering starts to dominate should a linear (bad metal) temperature dependence reappear (this regime, typically well above room temperature, has not been studied in most heavy fermion compounds).  In any case, the linear-in-temperature strange metal resistivity of heavy fermion compounds is only seen at much lower temperatures and resistivities, for YbRh$_2$Si$_2$ below 15\,K and 30\,$\mu\Omega$cm \cite{Tro00.2}.

\section{Strange metal behavior and Fermi surface jumps}\label{FSjumps}
In Section\;\ref{planckian}, a simple Drude form was used for the electrical resistivity and all temperature dependence was attributed to the scattering rate. Then, the question was asked which quasiparticles (with which $m/n$) to take if the scattering were to be Planckian. The answer was that this would have to be very weakly interacting quasiparticles, certainly not the ones close to the QCP from which the strange metal behavior emerges. Here we address another phenomenon that may challenge a Planckian scattering rate picture: Fermi surface jumps across these QCPs.

\begin{figure}[t!]
\hspace{0cm}\includegraphics[width=0.7\textwidth]{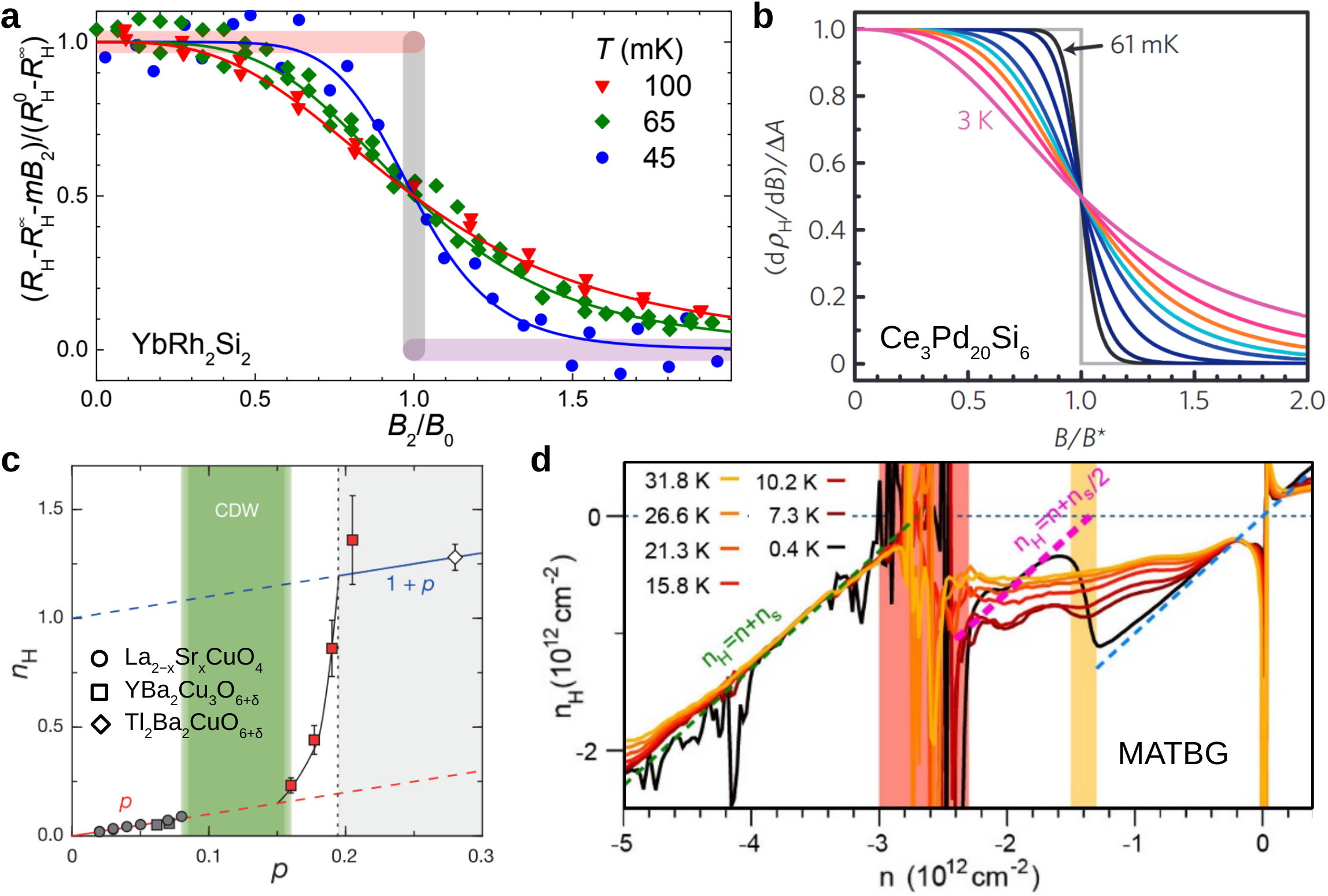}
\caption{Fermi surface jumps as evidenced by Hall effect measurements in several strange metals. (\textbf{a}) YbRh$_2$Si$_2$. From \cite{Pas21.1,Fri10.2}. (\textbf{b}) Ce$_3$Pd$_{20}$Si$_6$. From \cite{Cus12.1}. (\textbf{c}) Substitution series of three high-$T_{\rm c}$ cuprates. From \cite{Bad16.1}. (\textbf{d}) MATBG. From \cite{Cao18.1}.
\label{fig5}}
\end{figure}

This phenomenon was first detected by Hall effect measurements on YbRh$_2$Si$_2$
\cite{Pas04.1,Fri10.2} (Figure\;\ref{fig5}a). Let us first recapitulate the
experimental evidence for a Fermi surface jump across a QCP, as put forward in
these works. Hall coefficient $R_{\rm H}$ (or Hall resistivity $\rho_{\rm H}$)
isotherms are measured as function of a tuning parameter $\delta$ (in case of
YbRh$_2$Si$_2$ the magnetic field) across the QCP. A phenomenological crossover
function, $R_{\rm H}^{\infty} - (R_{\rm H}^{\infty}-R_{\rm
H}^0)/[1+(\delta/\delta_0)^p]$ \cite{Pas04.1}, is fitted to $R_{\rm H}(\delta)$
[or to $d\rho_{\rm H}/d B(\delta)$] and its full width at half maximum (FWHM) is
determined as a reliable measure of the crossover width. Only if this width
extrapolates to zero in the zero-temperature limit a Hall coefficient jump is
established. Of course, the jump size must remain finite in the zero temperature
limit. To identify a Fermi surface jump, other origins of Hall effect changes
must be ruled out, for instance anomalous Hall contributions from abrupt
magnetization changes at a metamagnetic/first order transition \cite{Chu09.1}.
All this was done for YbRh$_2$Si$_2$ \cite{Pas04.1,Fri10.2}. For
Ce$_3$Pd$_{20}$Si$_6$, using a very similar procedure, two Fermi surface jumps
were found at the two consecutive QCPs (Figure\;\ref{fig1}d)
\cite{Cus12.1,Mar19.1}. The crossover at the first QCP \cite{Cus12.1} is shown
in Figure\;\ref{fig5}b. It is also important to remind oneself that no Fermi
surface discontinuity is expected at a conventional antiferromagnetic QCP as
described by the spin density wave/order parameter scenario \cite{Col01.1}. Band
folding of the (even at $T=0$) continuously onsetting order parameter can in
that case only lead to a continuously varying Hall coefficient, as seen for
instance in the itinerant antiferromagnet Cr upon the suppression of the order
by doping or pressure (see \cite{Si13.1} for more details and the original
references).


These jumps are understood as defining signatures of a Kondo destruction QCP,
first proposed theoretically \cite{Si01.1,Col01.1,Sen04.1} in conjunction with
inelastic neutron scattering experiments on CeCu$_{5.9}$Au$_{0.1}$
\cite{Sch00.1}. At such a QCP, the heavy quasiparticles, composites with $f$ and
conduction electron components, disintegrate. The Fermi surface jumps because
the local moment, which is part of the Fermi surface in the paramagnetic Kondo
coherent ground state \cite{Osh00.1}, drops out as the $f$ electrons localize.
As such, Kondo destruction QCPs are sometimes referred to as $f$-orbital
selective Mott transitions. More recently, THz time-domain transmission
experiments on YbRh$_2$Si$_2$ thin films grown by molecular beam epitaxy
revealed dynamical scaling of the optical conductivity \cite{Pro20.1}. This
shows that the charge carriers are an integral part of the quantum criticality,
and should not be seen as a conserved quantity that merely undergo strong
scattering (as in order-parameter descriptions with intact quasiparticles). We
also note that a Drude description of the optical conductivity fails rather
drastically in the quantum critical regime \cite{Pro20.1}. It is thus unclear
how this physics could be captured by the simple Planckian scattering approach
described above.

Interestingly, Hall effect experiments in other strange metal platforms also
hint at Fermi surface reconstructions. Two examples are included in
Figure\;\ref{fig5}: a series of substituted high-$T_{\rm c}$ cuprates
\cite{Bad16.1} (panel c) and MATBG as function of the total charge density
induced by the gate \cite{Cao18.1} (panel d). Evidence for related physics has
also been found in the pnictides \cite{Yi15.1}. The physics here appears to be related to the presence of $d$ orbitals with a different degree of localization, with one of them undergoing a Mott transition, such as described by multi-orbital Hubbard models \cite{Yu17.1,Kom17.1}. It may well be that Fermi surface jumps are an integral part of strange metal physics, and should be included as a starting point in its description.

\section{Summary and outlook}
We have revisited the question whether the strange metal behavior encountered in numerous strongly correlated electron materials may be the result of Planckian dissipation. For this purpose, we have examined strange metal heavy fermion compounds. Their temperature--tuning parameter phase diagrams are particularly simple: Fans of strange metal behavior emerge from quantum critical points, in a Fermi liquid background. This, together with the extreme mass renormalizations found in these materials, makes them a particularly well-suited testbed.

As done previously, we use the Drude form of the electrical conductivity as a starting point, but complementary to a previous approach based on quantum oscillation data, we here rely on the Fermi liquid $A$ coefficient as precise measure of the quasiparticle renormalization. We find that for any of the measured $A$ coefficients, the slope of the linear-in-temperature strange metal resistivity $A'$ is much smaller than the value expected from Planckian dissipation. We also propose a new plot that allows to read off the ratio of effective mass to carrier concentration that one would have to attribute to the quasiparticles for their scattering to be Planckian. It corresponds to modest effective masses, perhaps something like a smooth background to quantum critical phenomena.

We have also pointed out that several heavy fermion compounds exhibit Fermi
surface jumps across strange metal quantum critical points and that this
challenges the Drude picture underlying the Planckian analysis. Indications for
such jumps are also seen in other platforms and may thus be a common feature of
strange metals. Further careful studies that evidence a sharp Fermi surface
change in the zero temperature limit, such as provides for some of the heavy
fermion compounds, are called for. On the theoretical side, approaches that
discuss the electrical resistivity as an entity and do not single out a
scattering rate as the only origin of strangeness are needed.

\vspace{6pt} 

\funding{This research has received funding from the European Union's Horizon 2020 Research and Innovation Programme under Grant Agreement no 824109 and from the Austrian Science Fund (FWF Grant 29296-N27).}

\acknowledgments{We acknowledge fruitful discussions with Joe Checkelsky, Piers Coleman, Pablo Jarillo-Herrero, Patrick Lee, Xinwei Li, Doug Natelson, T.\ Senthil, and Qimiao Si.}

\conflictsofinterest{The authors declare no conflict of interest.}

\end{paracol}

\reftitle{References}


\end{document}